\documentclass[%
  reprint,
superscriptaddress,
showpacs,preprintnumbers,
nofootinbib,
 amsmath,amssymb, 
 aps,
 prd
]{revtex4-2}
\pdfoutput=1 
\usepackage{amsmath,amstext,amssymb}
\usepackage[T1]{fontenc}
\usepackage{apjfonts} 
\usepackage{subfigure}
\usepackage{xcolor}
\usepackage{soul}
\usepackage{comment}
\usepackage{graphicx}
\usepackage[normalem]{ulem}
\usepackage[export]{adjustbox}
\usepackage[colorlinks=true,allcolors=blue]{hyperref}
\usepackage{orcidlink}

\graphicspath{{figures/}}

\newcommand{\de}{{\rm d}}

\def\apjl{\rm ApJL}
\def\apjs{\rm ApJS}

\def\aap{\rm A\&A}

\begin{document}

\title{A standard siren measurement of the Hubble constant using GW170817 and the latest observations of the electromagnetic counterpart afterglow}

\author{A.~Palmese~\orcidlink{https://orcid.org/0000-0002-6011-0530}}\thanks{NASA Einstein Fellow}\email{palmese@cmu.edu}
\affiliation{Department of Physics, University of California Berkeley, 366 LeConte Hall MC 7300, Berkeley, CA, 94720, USA}
\affiliation{McWilliams Center for Cosmology, Department of Physics, Carnegie Mellon University, Pittsburgh, PA 15213, USA}

\author{R.~Kaur}
\affiliation{Department of Physics, University of California Berkeley, 366 LeConte Hall MC 7300, Berkeley, CA, 94720, USA}
\affiliation{Department of Astronomy, University of California Berkeley, 501 Campbell Hall 3411, Berkeley, CA, 94720-3411, USA}

\author{A.~Hajela~\orcidlink{https://orcid.org/0000-0003-2349-101X}}
\affiliation{DARK, Niels Bohr Institute, University of Copenhagen, Jagtvej 128, 2200 Copenhagen, Denmark}

\author{R.~Margutti}
\affiliation{Department of Astronomy, University of California Berkeley, 501 Campbell Hall 3411, Berkeley, CA, 94720-3411, USA}


\author{A.~McDowell}
\affiliation{Center for Cosmology and Particle Physics, Physics Department, New York University, New York, NY 10003, USA}

\author{A.~MacFadyen}
\affiliation{Center for Cosmology and Particle Physics, Physics Department, New York University, New York, NY 10003, USA}

\begin{abstract}
    We present a new constraint on the Hubble constant $H_0$ using the latest measurements of the electromagnetic counterpart to the gravitational wave (GW) event GW170817. We use the latest optical, X-ray and radio observations of  the afterglow up to $\sim 3.5$ years after the GW detection, and properly take into account the impact of the host galaxy peculiar velocity. We find $75.46^{+5.34}_{-5.39}$ km s$^{-1}$ Mpc$^{-1}$ (68\% Credible Interval), a $\sim7\%$ precision measurement, which is a significant improvement compared to the 14\% precision of the first standard siren measurement. Our result is consistent within $1\sigma$ with the Cepheid--anchored Supernova and within $1.5 \sigma$ with the Cosmic Microwave Background measurements of the Hubble constant. We also explore the impact of the various assumptions made when fitting for the afterglow on the Hubble constant estimate.
\end{abstract}


\maketitle

\section{Introduction}

The Hubble constant ($H_0$) is a cosmological parameter that describes the current rate of expansion of the Universe. There currently exist a wide range of cosmological probes used to measure its value, although a 4-6$\sigma$ discrepancy currently exists between the measurements from those methods \cite{riess2021}. By measuring the cosmic microwave background (CMB) anisotropies, the Planck Collaboration \cite{planck18} inferred a value of $H_0 = 67.36 \pm 0.54$ km s $^{-1}$ Mpc $^{-1}$ under a $\Lambda$CDM cosmology. In contrast, through \emph{Hubble Space Telescope} observations, Cepheid variable stars and type Ia supernovae (SNeIa) provide a local Universe measurement of $H_0 = 73.04\pm1.04$ km s$^{-1}$ Mpc$^{-1}$ \cite{Riess_2022}. Currently, it is unclear where this Hubble constant tension arises from, and there is no agreement on whether it is caused by beyond--$\Lambda$CDM physics, or whether it is some source of systematics that is currently unknown \cite{2017NatAs...1E.169F,Di_Valentino_2021,cosmology_intertwined}.  

Gravitational wave (GW) measurements may help to clarify the origin of this discrepancy, as they can provide a measurement of $H_0$ which is independent of the set of systematics that may affect the aforementioned cosmological probes. Gravitational wave detection is currently possible thanks to the Advanced LIGO (Laser Interferometer Gravitational-Wave Observatory; \cite{2015LIGO}) and Virgo detectors \cite{Acernese_2014}. From the GW strain signal, one can infer the luminosity distance to the source of the event, so that GW detections can be used as ``Standard Sirens'' \cite{schutz}. Unlike SNeIa, for which it is often trivial to pinpoint the host galaxy and measure its redshift, gravitational wave detections alone do not usually allow a precise redshift measurement nor the precise location of the host galaxy of the source, so that has to be measured through another method. One way to obtain the redshift of a gravitational wave source is to pinpoint the host galaxy by finding its electromagnetic (EM) counterpart, and then fit the distance-redshift relation to derive the Hubble parameter through what we call the ``bright'' standard siren method. If no EM counterpart is identified, it is still possible to derive the redshift to the source in other ways through the so called ``dark'' standard siren methods. First, as originally proposed in \cite{schutz}, one can take advantage of galaxy catalogs to infer the source redshift by taking into account all of the potential host galaxies' redshifts through a dark standard siren galaxy catalog approach \cite[e.g.][]{delpozzo,darksiren1,fishbach,palmese20_sts,Gray_2020,Palmese_2021,Hitchhiker}. Alternatively, one can adopt a purely GW approach by making assumptions about the astrophysical mass distribution of black holes or neutron stars, and by doing so break the degeneracy between mass and redshift in the redshifted mass measured with the GW data \cite[e.g.][]{1993ApJ...411L...5C,Taylor_2012,Farr_2019}. While bright standard sirens are significantly rarer than dark standard sirens, they provide a significantly more precise Hubble constant measurement than the latter on an event-to-event basis \cite{chen17}, which is why they are powerful probes to understand the Hubble tension.

 The only gravitational wave event to date to be accompanied by an EM counterpart is the binary neutron star merger GW170817 \cite{ligobns}, which was detected on August 17, 2017 at 12:41:04 (UT) by the Advanced LIGO and Advanced Virgo detectors. The EM counterpart was discovered by several teams at different wavelengths \cite{MMApaper}. During the search for the counterpart in different wavelengths, a transient called ``kilonova'' associated with the GW event was discovered in the optical \cite[e.g.][]{Coulter1556,Soares_Santos_2017}, which enabled the identification of the host galaxy of the event: NGC 4993 \cite[e.g.][]{2017ApJ...849L..34P}. As this was the first GW multi-messenger observation, the first standard siren measurement of the distance to the source of the event was derived using the GW data and the host galaxy redshift \cite{2017Natur.551...85A}. 

One limitation with measuring distances from gravitational waves, especially considering current generation ground-based GW detectors, is that the distance measurement is degenerate with respect to the inclination angle of the system. This impacts how precisely distances can be measured from GW data alone. On the other hand, if a gravitational wave event has an EM counterpart, we gain more insight on the properties of the event, including its extrinsic, geometrical properties.

Binary neutron-star mergers, such as the one from GW170817, may result in the launch of relativistic outflows (e.g. jets) that interact with the ambient medium to produce broadband synchrotron emission \cite{2020PhR...886....1N,doi:10.1146/annurev-astro-112420-030742}, called ``afterglow''. 
In scenarios where jets are not aligned with our line-of-sight (viewed off-axis), the observed time of peak emission depends on the geometry of the outflow (mostly on the parameters viewing angle $\theta_{\rm obs}$ and jet opening angle $\theta_{\rm jet}$ \cite{Nakar_2021}). The emission from GW170817 has been dominated by the afterglow from an off-axis structured jet, at least on timescales beyond that of the GW, $\gamma$-ray burst, and the thermal kilonova emission; here we use the EM data from the observations of this jet afterglow.

In this paper, we derive an improved measurement of $H_0$ by deriving updated viewing angle constraints from the EM counterpart using data out to over 3 years from the GW merger, in place of the data up to $\sim 1$ year, state-of-the-art hydrodynamic simulations from \cite{mcdowell}, as well as including a proper treatment of the impact of peculiar velocity of the host galaxy \cite{nicolaou2019impact}. Compared to other works that have previously measured the Hubble constant with GW170817 and viewing angle constraints, we use more recent afterglow data over a larger set of wavelengths (e.g. compared to \cite{Guidorzi:2017ogy,Wang21}) with a hierarchical Bayesian formalism from the afterglow fitting to the $H_0$ estimate, while also taking into account the impact of assumptions on the jet modeling and the uncertainty contribution that arises from a proper treatment of the host galaxy's peculiar velocity. Compared to  \cite{Hotokezaka,Mukherjee}, we focus on the afterglow rather than the jet superluminal motion. For the peculiar velocity treatment, this is also done in different ways in \cite{nicolaou2019impact,Mukherjee,howlett_davis}. See \cite{Bulla22} for a review and discussion of existing works on this topic.

We can expect an improved constraint compared to using previous afterglow observations as  observations further beyond the peak flux are now available. The width of the peak does provide some information about the ratio of the jet opening angle and the viewing angle \cite{Nakar_2021}. Given the current brightness of the EM emission from GW170817 \cite{Hajela:2021faz} it is unlikely that a significant amount of new detections will be obtained in the future (unless a re-brightening occurs). This work thus aims to provide a comprehensive measurement of the Hubble constant from GW170817 and the electromagnetic counterpart.

In Section \hyperref[section:Data]{2} we discuss our GW and EM data. In Section \hyperref[section:Methods]{3} we discuss the derivation of the Hubble constant posterior, and how we measure the peculiar velocity and the viewing angle. In Section \hyperref[section:Results]{4} we discuss our findings, and explain how they compare to earlier works that derived $H_0$ with GW170817. Lastly, in Section \hyperref[section:Conclusion]{5}, we state our final conclusions from this work and possible future steps.

\section{Data}
\label{section:Data}

\cite{Hajela:2021faz} present the latest afterglow observations made with \emph{Chandra X-ray Observatory (\textit{CXO})}, Karl G. Jansky Very Large Array (VLA), \emph{Hubble Space Telescope (\textit{HST})}, and MeerKAT. We use their data out to $\sim 4$ years after the event in the X-ray and radio wavelengths. We do not include the Australian Telescope Compact Array (ATCA) data, as they suffer from a systematic calibration issue and do not add more information compared to VLA. Similarly, we ignore \textit{XMM}-Newton data as the PSF is much larger than Chandra, and as a result they also suffer from a systematic offset compared to the latter, without adding constraining power to our analysis. The data used is from a compilation of a number of works \cite{Hallinan_2017,alexander_2017,Mooley_2017,margutti_2017,Margutti_2018,Dobie_2018,Alexander_2018,mooley,Hajela_2019,Fong2019}. We do not use the same compilation of afterglow observations \cite{Makhathini_2021} considered in the viewing angle estimates from the jet superluminal motion \cite{Mooley_2022}, as a spectral photon index is assumed to calibrate the late-time observations in that data set (while we are fitting this value across all data points in the current analysis). See \cite{Hajela:2021faz} for a detailed discussion of the data analysis and the differences with respect to the reduction in the two studies.

In addition to the EM data, the GW data we use in our fiducial analysis are the high-spin samples from LIGO/Virgo \cite{abbott2019properties} at \url{https://dcc.ligo.org/LIGO-P1800061/public}.

 We use the catalog of peculiar velocity data from the 6dF Galaxy Survey \cite{springob20146df} to calculate the weighted peculiar velocity of NGC 4993.

\section{Method}\label{section:Methods}

\subsection{Hubble constant posterior}
\label{section:Hubble}

In this subsection, we derive the Hubble constant posterior, taking into account both the viewing angle measurements from the afterglow and a peculiar velocity treatment that marginalizes over the smoothing scale as in \cite{nicolaou2019impact}.
Similarly to \cite{2017Natur.551...85A}, we write the likelihood of observing the GW and the EM data  $x=(x_{\rm GW}, x_{\rm EM})$, and of measuring a recessional velocity $v_r$ for the host galaxy and a peculiar velocity field $\langle v_p \rangle$ at the location of the host as:

\begin{equation}
\begin{split}
    & p(x,v_r, \langle v_p \rangle|d, \cos{\iota}, v_p, H_0)  = \\
    & =p(x_{\rm GW}|d, \cos{\iota})p(x_{\rm EM},v_r|d, \cos{\iota}, v_p, H_0) p(\langle v_p \rangle|v_p)\\
    &=p(x_{\rm GW}|d, \cos{\iota})p(x_{\rm EM}|d, \cos{\iota}, v_r) p(v_r|d,  v_p, H_0)p(\langle v_p \rangle|v_p)\, ,
\end{split}
\end{equation}
where $d$ is the luminosity distance, $\iota$ is the inclination angle, and $v_p$ is the peculiar velocity of the host.
From the marginal likelihood we derive the posterior on $H_0$, after marginalizing over $d, \cos{\iota}, v_p$, and including the marginalization over the smoothing kernel as in \cite{nicolaou2019impact}:

\begin{widetext}
\begin{equation}
    p(H_0 | x,v_r, \langle v_p \rangle)  
     \propto \frac{p(H_0)}{\mathcal{N}_s(H_0)}~\int \de d~ \de v_p ~\de \cos{\iota} ~\de s ~p(x_{\rm GW}|d, \cos{\iota}) ~p(x_{\rm EM}|d, \cos{\iota}, z(v_r)) ~p(v_r|d,  v_p, H_0)~p(\langle v_p \rangle|v_p,s)
     ~p(d)~p(\cos{\iota})~p(v_p)~p(s),
\end{equation}\label{eq:posterior}
\end{widetext}

\begin{figure*}[ht]
    \centering
    \includegraphics[width=\textwidth]{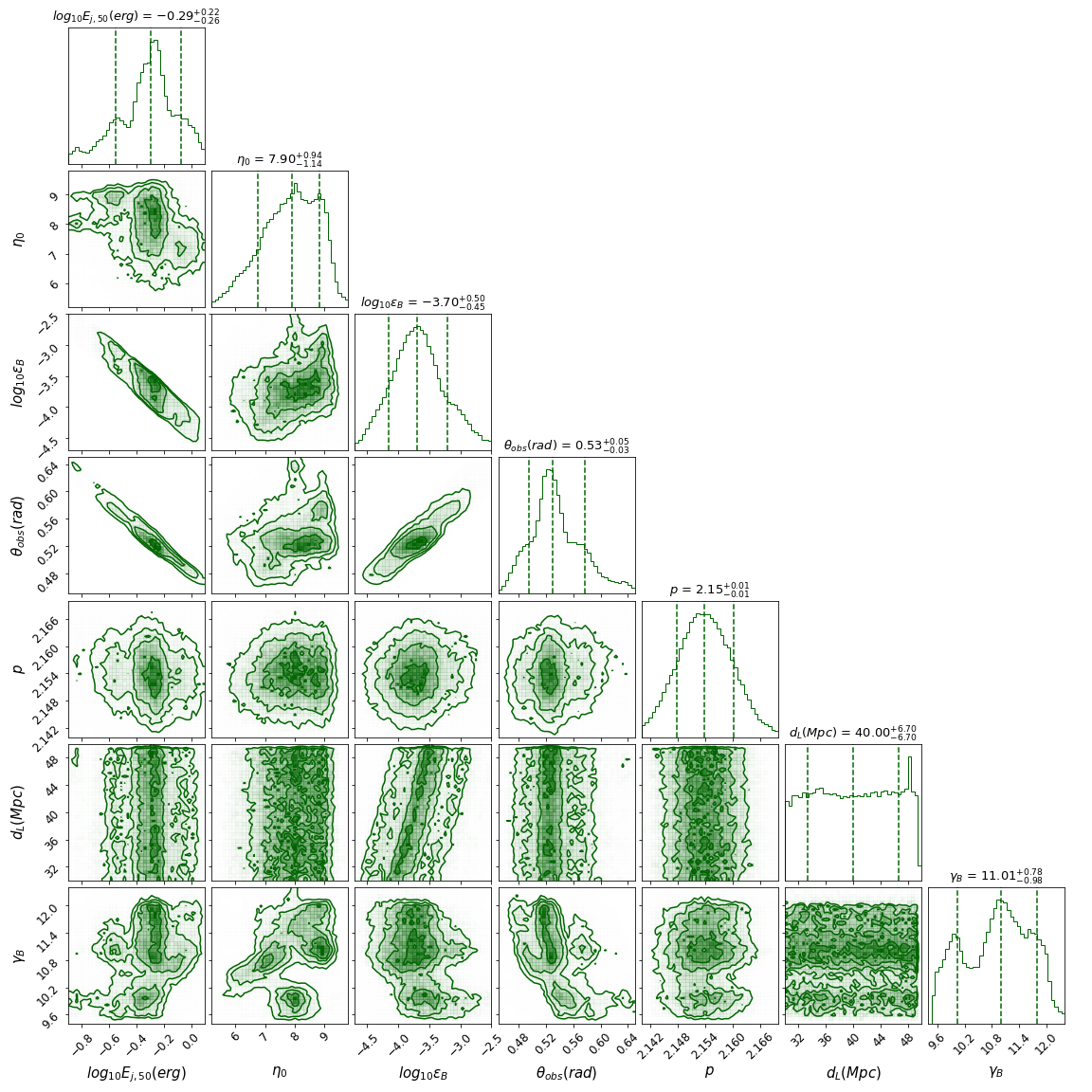}
    \caption{\texttt{JetFit} corner plot: 1D and 2D projections of the parameters we fit using the afterglow light curve. Vertical dashed lines mark the 16th, 50th, and 84th percentiles of the distributions.
    }\label{fig:jetfit}
\end{figure*}

where the likelihood of the recessional velocity is a Gaussian as in \cite{2017Natur.551...85A}: $p(v_r|d,  v_p, H_0)=\mathcal{N}(v_p+H_0d, \sigma_{v_r}; v_r)$, where $v_r =3327$ km s$^{-1}$, and $\sigma_{v_r} = 72$ km s$^{-1}$. The term $~p(x_{\rm EM}|d, \cos{\iota}, z(v_r))$ is the likelihood of the EM data, and we recover it from the \texttt{JetFit} \cite{Wu_2018} output samples. The term $p(x_{\rm GW}|d, \cos{\iota})~p(d)~p(\cos{\iota})$ is effectively the posterior from the GW data that we get from the publicly available release. 

The peculiar velocity  term is a Gaussian as in \cite{nicolaou2019impact}: $~p(\langle v_p \rangle|v_p,s)=\mathcal{N}[v_p, \sigma_{v_p}](\langle v_p \rangle(s))$ with the best fitting values of $v_p$ and the smoothing scale $s$ derived as follows, similarly to \cite{nicolaou2019impact}. From the 6dF Galaxy Survey (6dFGSv) \cite{springob20146df}, we calculate the peculiar velocities ($v_p$) of the galaxies surrounding NGC 4993, the host galaxy of the binary neutron star merger GW170817. Following \cite{springob20146df}, we use:

\begin{equation}
    v_p = cz(1-10^{-\delta})
\end{equation}

where $z$ is the CMB frame galaxy redshift, and $\delta$ is the logarithmic distance ratio $D_z/D_H$, where $D_z$ and $D_H$ are the linear comoving distances of the galaxy corresponding to the observed redshift and Hubble redshift, respectively. For the complete derivation of this equation, see \cite{springob20146df}. To obtain the peculiar velocity of the host galaxy of the merger, we weigh the peculiar velocities of the other galaxies in the catalog according to their distance from NGC 4993 by using a 3D Gaussian kernel centered the location of NGC 4993. As changing the width of the kernel affects the weighted peculiar velocity that we get, which in turn affects the value of $H_0$ that we calculate, we call the width of the kernel the smoothing scale $s$, include it in our likelihood function, and marginalize over it.

For the prior on $s$ we follow \cite{nicolaou2019impact} and use a Gamma distribution with shape 2 and scale 4 h$^{-1}$ Mpc: $p(s) \propto \Gamma[2, 4]$. We use a uniform prior on $v_p$ such that: $p(v_p) \propto U[-1000, 1000]$ km s$^{-1}$. The prior on $H_0$ is flat in log, consistent with most of the other $H_0$ estimations from GW170817.

At last, the term $\mathcal{N}_s(H_0)$ is a normalization factor that takes into account the selection effects that may arise from the GW and EM observations. Selection effects are not deemed significant for GW170817 \cite{2017Natur.551...85A,Mastrogiovanni_2021}, hence we ignore them in this work.
We sample the posterior using a Markov-Chain Monte Carlo (MCMC) sampling, namely through the package \texttt{emcee} \cite{Foreman_Mackey_2013}.

\subsection{Viewing angle constraints}
\label{section:View}

We use the EM data and the package \texttt{JetFit} \cite{Wu_2018} to fit for the jet parameters, including the viewing angle, which is used in our Hubble posterior to break the distance-inclination angle degeneracy. The advantage of using this code is that it is based on hydrodynamic simulations together with an analytical and physically motivated boosted fireball model for structured jets \cite{Duffell13} able to reproduce a variety of outflow structures. This feature allows us to probe the effect of various assumptions on the input parameters and outflow structures on the viewing angle estimates.


The synchrotron emission calculated by \texttt{JetFit} depends on the following parameters: the explosion energy, $E_0$; the ambient density, $n$; the specific internal Energy of the fireball, $\eta_0 \simeq E/M$; the bulk Lorentz factor, $\gamma_B$; the spectral index of the electron distribution, $p$;  the fraction of energy that goes into shocked electrons, $\epsilon_e$;  the fraction of energy that goes into magnetic energy, $\epsilon_B$, and the viewing angle, $\theta_{\rm obs}$. We derive posterior distributions on these parameters using \texttt{JetFit}.
We fix $n$ to 0.01 cm$^{-3}$ (close to the limit found in \cite{Hajela_2019}) and do not explore the effect of varying this parameter, since any variation in $n$ is degenerate with $E_0$. 
The synchrotron emission of GW170817 is completely determined by the ratio $E/n$, and we will not attempt to estimate these two parameters independently as this is not relevant for our analysis (as also done in \cite{Hotokezaka}). 
We consider several priors for $\gamma_B$, including fixing it to 12, as radio observations of the jet motion suggest a very small $\lesssim 5^o$ jet opening angle \cite{mooley,ghirlanda}, corresponding to $\gamma_B\simeq 1/ \theta_{\rm jet}\simeq 12$. We use the latest simulations from \cite{mcdowell}, which are updated compared to \cite{Wu_2018}, and allow for a wider range of $\gamma_B$ up to 20 (while \cite{Wu_2018} only goes to 12), which is important for the case of GW170817 given the potentially small opening angle.


We test letting $\epsilon_e$ free, and we find that it has no effect on $\theta_{\rm obs}$ or $d_L$, so we fix $\epsilon_e$ to 0.1 (as predicted from the simulations of particle acceleration by relativistic shocks \cite{sironi2013,Sironi_2015}). We use a uniform prior on $d_L$, $3\sigma$ around the distance from LIGO for our fiducial analysis (although this specific choice does not have a significant effect on the analysis, given the mild degeneracies of $d_L$ with other parameters). We fix the redshift to that measured from the recessional velocity of the host, and let all other parameters free. 

Since the GW data is in the form of inclination angle $\cos{\iota}$, and the \texttt{JetFit} sampling of the afterglow data gives us the viewing angle $\theta_{\rm obs}$ with respect to the jet, we must perform a conversion so that the same variable is considered in the posterior. Assuming that the jet is aligned with the binary angular momentum, the relation between the two is defined as: $\theta_{\rm obs}=$min$(\iota, 180^o-\iota)$. GW measurements allow for the identification of the type of rotation (clockwise or counter-clockwise), and as such, the inclination angle ranges from $\iota=0^o$ (counter-clockwise) to $\iota=180^o$ (clockwise). In contrast, EM data does not allow us to determine this type of rotation, so it only ranges from $\theta_{\rm obs}=0$ to $\theta_{\rm obs}=90^o$. In going from inclination angle to viewing angle, we lose information, so it is not possible to extrapolate $\iota$ given only $\theta_{\rm obs}$. 


We therefore convert the GW data to viewing angle, and use the Gaussian Mixture Model feature of \texttt{scikit-learn} to reconstruct the 2D posterior $p(d, \cos{\iota}|x_{\rm EM})$ from the MCMC samples of $\theta_{\rm obs}$ and $d_L$, so that we can compute the posterior of Eq. \hyperref[eq:posterior]{2} at the locations of our samples. For the Gaussian Mixture Model, we use 12 mixture components.



\begin{figure}[h]
    \center
    \includegraphics[width=\columnwidth]{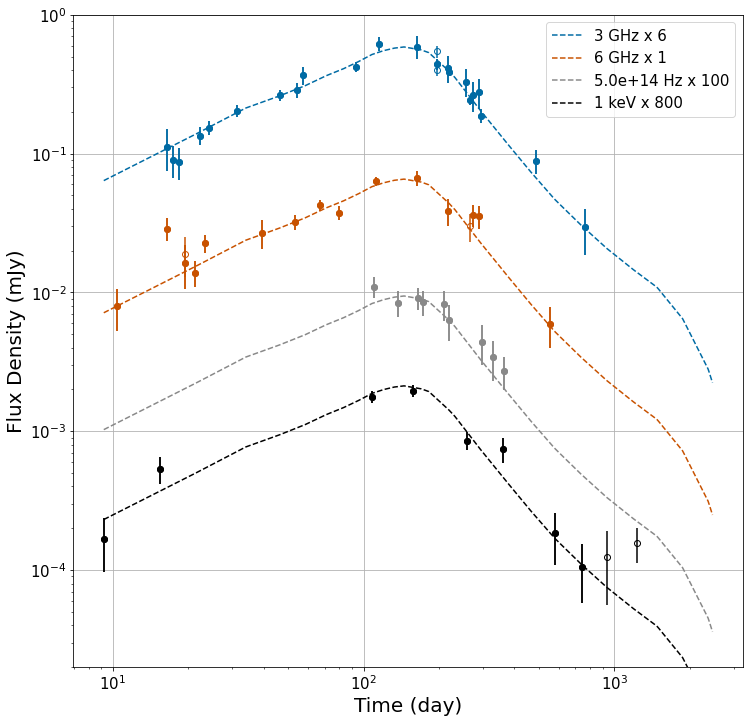}
    \caption{GW170817 radio to X-ray afterglow light curves used in this work, with the best fit from \texttt{JetFit} (dashed lines). Data is shifted by an arbitrary factor indicated in the legend for visualization purposes. The radio observations are in blue and orange, the X-ray observations from CXO in dark gray, and the UV/optical from \emph{HST} in light gray. Open circles indicate data that we have added compared to the fiducial fit of \cite{Hajela_2019}, including points at $t>900$ days after the merger.}\label{fig:lightcurves}
\end{figure}

\begin{figure}[h]
    \center
    \includegraphics[width=\linewidth]{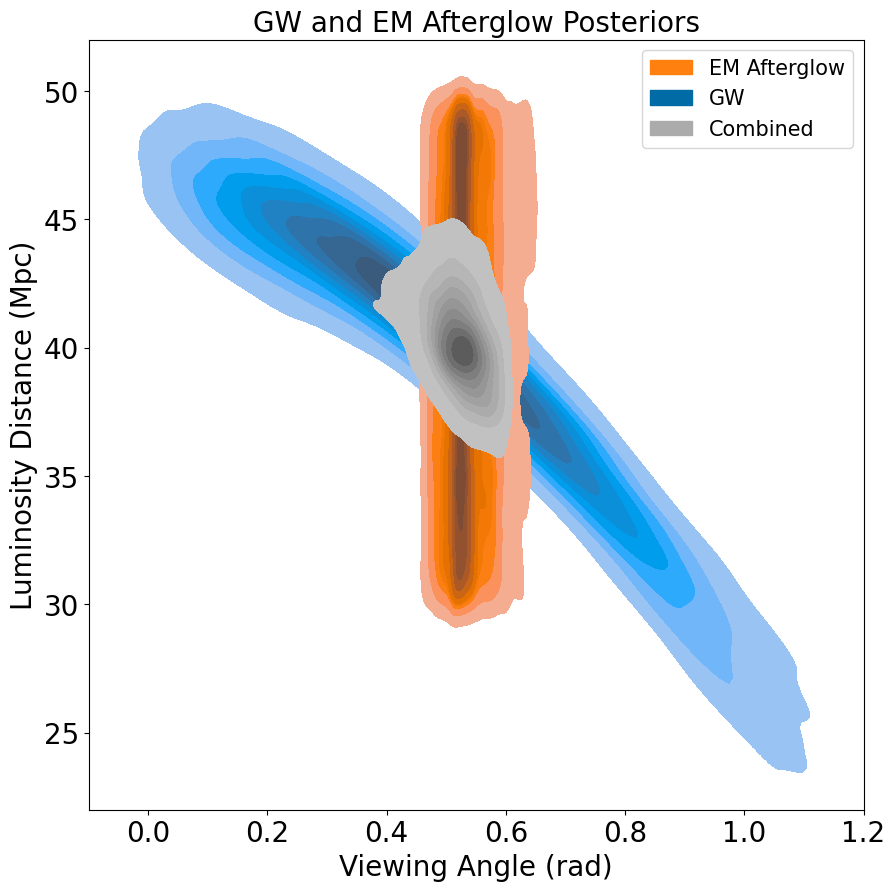}
    \caption{Joint luminosity distance and viewing angle posteriors from GW data alone (blue), the afterglow data (orange; with a broad uniform prior in distance), and from the joint GW and EM analysis (grey). The density levels go from 10 to 90\% credible interval.}\label{fig:SNS}
\end{figure}

\begin{figure*}[ht]
    \centering
    \includegraphics[width=0.8\textwidth]{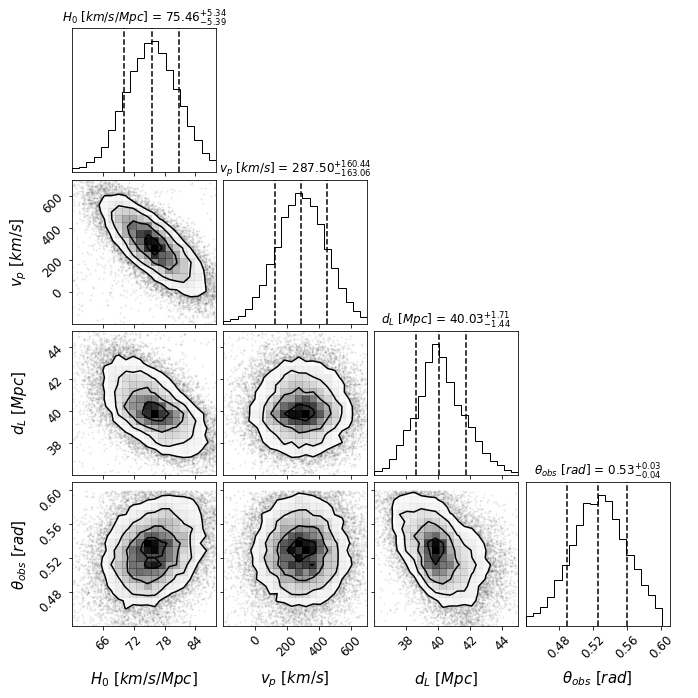}
    \caption{Corner plot showing the 2D posteriors for some parameters of interest (peculiar velocity, luminosity distance, and viewing angle of the binary) from our multi-messenger analysis. The strongest degeneracy here appears between $H_0$ and the peculiar velocity, showing that a careful treatment of the peculiar velocity field such as in \cite{nicolaou2019impact} is of major importance in this kind of analysis to avoid incurring in biased estimates.
    }\label{fig:H0corner}
\end{figure*}

\section{Results and discussion}

\begin{table*}
\centering
    \begin{tabular}{c c c c c c c c} 
    \hline
    \hline
     Reference  & $H_0$ [km/s/Mpc] & $\theta_{\rm obs}$ [deg]
     & $\theta_{\rm jet}$ [deg]& $d_L$ [Mpc] & EM data & Time & EM counterpart\\ 
     \hline
 \cite{2017Natur.551...85A} & $70.0^{+12.0}_{-8.0}$ & -- & --& $43.8^{+2.9}_{-6.9}$ & -- & -- & KN id\\  
 
 \cite{nicolaou2019impact} & $68.6^{+14.0}_{-8.5}$ & --& -- & -- & -- & -- & KN id \\

 \hline
 
 \cite{Guidorzi:2017ogy} & $75.5^{+11.6}_{-9.6}$ & 25-50 & 15 & $39.5$ & X-ray (\emph{Chandra}),  & $<40$ days& Afterglow\\
 &&&&&Radio (VLA)& &\\

 \cite{Hotokezaka} & $70.3^{+5.3}_{-5.0}$ & $15 < \theta_{\rm obs} \frac{d}{41~ \text{Mpc}}  < 29$
 & < 5.7 & 41 & Radio (VLBI) & 75-230 days & Jet  motion \\
 
  \cite{Mukherjee} & $68.3^{+4.6}_{-4.5}$ & -- & --  & -- & Radio (VLBI) & -- & Jet  motion \\
  
  \cite{Dhawan_2020} & $72.4^{+7.9}_{-7.3}$ & $32.5^{+11.7}_{-9.7}$ & -- & -- & UV, Optical, NIR & $< 10$ days & KN \\
  
 \cite{Wang21} & $69.5 \pm 4$ & $22\pm1$ & $\sim 11$ & $43.4\pm 1$ & Radio (VLA 3GHz) &$<$ 300 days& Afterglow\\

 This work & $75.46^{+5.34 }_{-5.39}$ & $30.4^{+2.9}_{-1.7}$ & $\sim 5 $& $41.7^{+1.5}_{-1.3}$&  X-ray (\emph{Chandra}), &$\lesssim $ 4 years & Afterglow \\
 &&&&& UV/Optical (\emph{HST}), & & \\
  &&&&& Radio (VLA) & \\
 
     \hline

    \end{tabular}
    \caption{Comparison of $H_0$ constraints and assumptions from various works discussed in this paper, including their estimate of the viewing angle and distance, where available. The last columns show which EM data and counterpart was used in each work to derive the related estimates. The top block does not include viewing angle constraints (hence it only uses the KN identification -- KN id -- in other words its location and host galaxy redshift), while the bottom does. For \cite{mukherjee2020standard} we do not include the jet motion constraints as those are the same as in \cite{Hotokezaka}.  }
\label{tab:H0}
\end{table*}

\begin{figure*}[ht!]
    \centering
    \includegraphics[width=170mm]{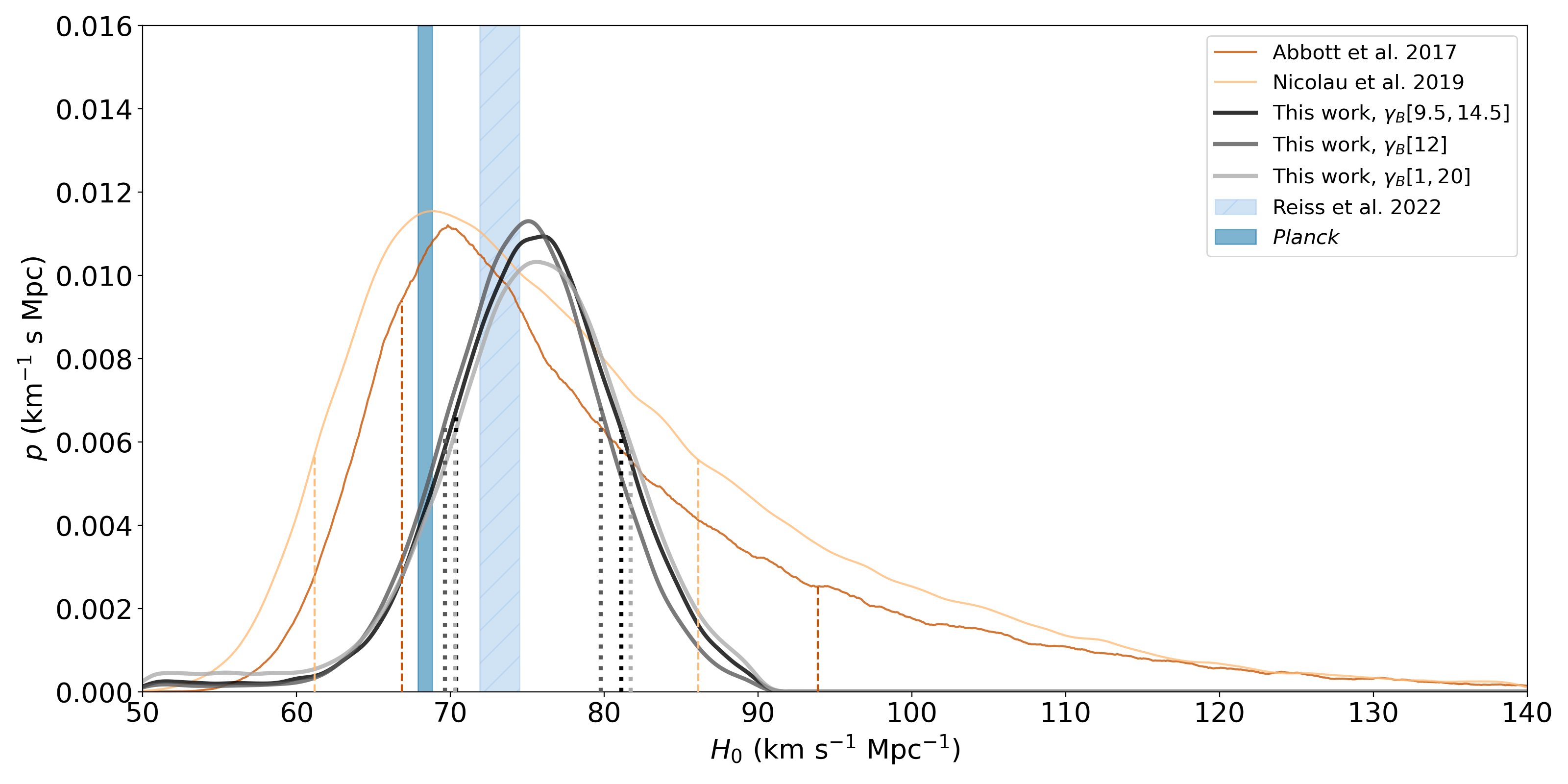}
    \caption{Plot of our resulting $H_0$ posterior distribution with the updated inclination angle constraints (black line) in comparison to the Planck (vertical dark blue band; \cite{planck18}), LVC and EM partners (dark orange line, not including EM viewing angle constraints; \cite{2017Natur.551...85A}), and SH0ES (vertical light blue band; \cite{riess2021}) constraints. To evaluate the improvement of the viewing angle estimates on $H_0$, our result needs to be compared to the posterior that does not include EM viewing angle information, but uses the same peculiar velocity treatment (light orange line; \cite{nicolaou2019impact}). The two grey posteriors show the impact of different $\gamma_B$ priors on $H_0$. The vertical dashed lines represent the lower and upper $1\sigma$ error bounds of each posterior.}\label{fig:posteriors_graph}
\end{figure*}

For our fiducial analysis with \texttt{JetFit}, using the parameters as described in section \hyperref[section:Methods]{2}, we find 
$\theta_{\rm obs}=0.53_{-0.03}^{+0.05}$ rad, and the constraints on all parameters are shown in Fig. \ref{fig:jetfit}. In this case, we use a prior in $\gamma_B$ which is uniform between 9.5 and 14.5, motivated by two reasons: 1) this corresponds to the range of $\sim4-6^o$ jet core opening angle found by the proper motion analysis of \cite{Mooley_2022}, and 2) the recovered posterior shows most of the support in this range, and tends to 0 above and below this range. Similarly to \cite{mcdowell}, when $n_0$ is fixed, we recover a $\gamma_B$ close to 11.

We see that some degeneracy is present, as expected, in the $\epsilon_B-E$ plane, since the observables depend on a combination of $E$, $\epsilon_B$, and $n$ (the latter of which we have fixed). We also note a degeneracy between energy and viewing angle, which therefore explains also a degeneracy between viewing angle and $\epsilon_B$. It is worth noting that also $\epsilon_B$ and distance are midly degenerate: we find that the effect of changing the former has a similar effect on the lightcurves to that of changing the distance, i.e. of shifting the curve towards higher or lower values, hence the observed degeneracy.  

The best fit lightcurves from this run are shown in Fig. \ref{fig:lightcurves}. We note that by including detections at more than 900 days after trigger, we may be including a rebrightening of the afterglow which may potentially be due to the kilonova ejecta afterglow (\cite{Hajela:2021faz}; see last data points in dark grey Fig. \ref{fig:lightcurves}). 
We test whether the inclusion of these late-time data has an effect on our analysis, and we find that including or excluding them in our fit does not result in any significant changes of our viewing angle constraint, so we decide to use those data for the purpose of this analysis. It is clear from Fig. \ref{fig:SNS}, showing joint luminosity distance-viewing angle posteriors, that the afterglow constraints (orange) are able to optimally break the distance-inclination angle degeneracy from the GW data alone (blue): the joint GW-EM posterior from this analysis is shown by the grey contours.
The fiducial constraint on the Hubble constant that we derive using this \texttt{JetFit} result  is $H_0=75.46^{+5.34}_{-5.39}$ km s$^{-1}$ Mpc$^{-1}$ (68\% Credible Interval). The posterior corner plot for the Hubble constant, the peculiar velocity, and the luminosity distance is shown in Fig. \ref{fig:H0corner}. A comparison plot of our $H_0$ posterior with other relevant measurements is shown in Fig. \ref{fig:posteriors_graph}, while Table \ref{tab:H0} summarizes several other results. Our work provides a significant improvement in precision compared to the original GW170817 standard siren measurement of \cite{2017Natur.551...85A}. 
While our result appears to be closer to the late-time Universe measurements of $H_0$ from Supernovae and Cepheids, we note that it is only $< 1.5\sigma$ away from the \emph{Planck} measurement, hence there is no discrepancy at the current level of precision.

We explore the impact of a range of assumptions made in the afterglow fit on the viewing angle constraints (so, indirectly, on the Hubble constant). We test that variations of $\epsilon_e$ do not have a significant impact on the viewing angle, and this parameter is well-known from shocks simulations. 
We find that the \texttt{Jetfit} viewing angle output is not significantly sensitive to the luminosity distance prior as fixing it to the mode of the GW posterior or leaving it free to vary within $3\sigma$ of the GW distance posterior does not change the viewing angle posterior significantly. In other words, these parameters are not significantly degenerate in the light curve fitting, at least as long as the redshift is fixed, while the degeneracy is strong in the GW posterior. The redshift is not from the Hubble expansion alone in this case and it is measured independently of the GW posterior, so it is safe to evaluate separately the GW and EM posteriors.

As expected, we find that the parameter that has the largest effect on the viewing angle is $\gamma_B$.  This is expected because $\gamma_B$ is directly related to the jet opening angle, which in turn is the parameter which is most degenerate with the viewing angle in the afterglow lightcurve fit. Works have however shown that the jet opening angle can be as small as $\lesssim 5^o$ \cite{mooley,ghirlanda,Mooley_2022}, corresponding to $\gamma_B\simeq 1/ \theta_{\rm jet}\gtrsim 11$, so excluding the region at $\gamma_B<9.5$ for our fiducial analysis is a reasonable assumption to explore. By opening up the $\gamma_B$ prior over the entire available range [1,20], the fit prefers lower values close to 8 (as also found in \cite{mcdowell}), which would be inconsistent with the superluminal motion measurements. On the other hand, $\gamma_B>14.5$ do not provide a good fit to the afterglow data. We conclude that our fiducial prior on $\gamma_B$ is able to provide constraints that are both consistent with the jet motion and the afterglow lightcurve. In other words, a $\gamma_B$ on the low-end side of that expected from \cite{mooley}, coupled with a viewing angle that is on the higher end, can reconcile the VLBI and afterglow measurements. Note that the effect of a varying $\gamma_B$ prior are minimal on the $H_0$ posterior (grey and black curves in Fig. \ref{fig:posteriors_graph}), although there is a shift in viewing angle posterior mode due to the aforementioned degeneracy (e.g $\theta_{\rm obs}=0.66^{+0.04}_{-0.11}$ rad for the widest $\gamma_B$ prior). This is because by marginalizing over $\gamma_B$ we are taking into account the existing $\gamma_B-\theta_{\rm obs}$ degeneracy and the uncertainties are large enough that the various viewing angle estimates turn out to be consistent with each other. The resulting shift in $H_0$ is at maximum $\sim 0.9$ km/s/Mpc, so that while this can be ignored at the current level of precision, future percent level standard siren measurements will have to take into account such potential systematic bias.  

We compare our result to what one would find using the constraints of \cite{Mooley_2022} from the jet motion. They find that the viewing angle is constrained to 19–25 deg at 90\% CI from optical superluminal jet motion and afterglow observations, while we find $30.4^{+2.9}_{-1.7}$ deg (68\% CI). A viewing angle of 19–25 deg does not seem to fit the afterglow late time evolution well (see e.g. Fig. 2 of \cite{Mooley_2022}), and we are unable to make the \texttt{JetFit} posterior chain converge when using a viewing angle prior of 19–25 deg and a $\gamma_B$ range consistent with the jet opening angle constraint from \cite{Mooley_2022}. As a further test of the impact of different viewing angle estimates on $H_0$, we assume a viewing angle constraint from \cite{Mooley_2022} approximated with a Gaussian (since we do not have access to the full posterior and are unable to reproduce it), and find $H_0 =71.93^{+4.29}_{-5.22}$ km/s/Mpc. Note that this specific constraint on the viewing angle assumes a luminosity distance, and that this test is only to understand the impact on $H_0$ of different viewing angle constraints, and it should not be taken as a fiducial value for the Hubble constant given the simplifications made. Under some simplifying assumptions \cite{Mooley_2022} find $H_0=71.5\pm 4.6 $ km/s/Mpc, which is broadly consistent with our findings.

Next, we compare our results to other works in the literature. Compared to \cite{Wang21} we use more recent afterglow observations, we jointly fit X-ray, optical, and radio data instead of radio alone, and add a peculiar velocity treatment. Although the latter adds to the error contribution of our final estimate, thanks to the added data we find a comparable 68\% interval of $H_0$. We also note that \cite{Wang21} and \cite{Hotokezaka} fix $p$ and the redshift to the cosmological redshift 
while we let $z$ vary according to our priors. 

It is worth noting that estimations of the inclination angle have been produced using the kilonova (KN) light curve and spectroscopy associated to GW170817 \cite{Dhawan_2020,perezgarcia}, however the modeling of KN lightcurves and its viewing angle dependence ha significant systematic biases and uncertainties \cite{2021MNRAS.502.3057H}, due also to the lack of KN observations so far, so we do not include KN constraints in this work.

Following \cite{nicolaou2019impact}, we note that our approach to the peculiar velocity treatment is more conservative with respect to \cite{Mukherjee}, thus a larger uncertainty contribution for $H_0$ is expected from peculiar velocities in our case.


\section{Conclusion}
\label{section:Conclusion}
 In this paper, we have shown how long-term, multi-wavelength monitoring of the jet afterglow that followed GW170817 can provide exquisite measurements of the viewing angle of the binary, which are in turn able to break the distance-angle degeneracy and improve standard siren constraints of the Hubble constant.
 Although \cite{Mastrogiovanni_2021} show that in the long run afterglow measurements will not be the main driver in the Hubble constant precision from standard sirens, \cite{Nakar_2021} claim that it is possible that a few events with precise viewing angle constraints will dominate the $H_0$ uncertainty from standard sirens. We conclude that, at least for nearby and close to on-axis events, EM viewing angle measurements can still provide a significant improvement to the Hubble constant constraints. We also find that percent level precision cosmology measurements from standard sirens taking advantage of afterglow viewing angle constraints will likely need a careful treatment of systematics from the modeling assumptions not to incur in significant biases.
 
 Future works should analyze realistic simulations of GW detections, including a possible LIGO Voyager and Next Generation (XG) GW detector networks, along with upcoming X-ray, UV, optical, and radio EM instruments' sensitivity, to assess the impact of afterglows on GW cosmology. In this work, a prior consistent with the jet motion was used to break the jet opening angle-viewing angle degeneracy from the afterglow data alone, but our results proved to be robust to varying assumptions. Although jet motion measurements will become increasingly challenging as the GW detectors sensitivity continues to improve and the detection horizon moves further, even viewing angle upper limits can significantly improve GW distance measurements \cite{Chen_2019}, while in XG afterglow identification may become sufficient thanks to the more precise distance measurements, so it remains to be proven whether afterglows will eventually become a powerful tool for GW cosmology.

\acknowledgments
\noindent  We thank Constantina Nicolaou for help with the peculiar velocity estimation, and Clecio Bom, Ignacio Maga\~na Hernandez, Geoffrey Ryan for helpful discussion. A. Palmese acknowledges support for this work was provided by NASA through the NASA Hubble Fellowship grant HST-HF2-51488.001-A awarded by the Space Telescope Science Institute, which is operated by Association of Universities for Research in Astronomy, Inc., for NASA, under contract NAS5-26555. R. Kaur was supported by the Physics Innovators PI$^2$ program at the University of California, Berkeley.  

\bibliographystyle{yahapj}
\bibliography{references1}


\end{document}